# Anomalous Nodal Gap in a Doped Spin-1/2 Antiferromagnetic Mott Insulator

Yong Hu[1,2,#], Christopher Lane[3,#], Xiang Chen[4], Shuting Peng[1], Zeliang Sun[1], Makoto Hashimoto[5], Donghui Lu[5], Tao Wu[1], Robert S. Markiewicz[6,7], Xianhui Chen[1], Arun Bansil[6,7], Stephen D. Wilson[4] and Junfeng He[1,*]

[1]*Department of Physics, University of Science and Technology of China, Hefei, Anhui 230026, China*

[2]*School of Physics and Institute of Advanced Interdisciplinary Studies, Chongqing University, Chongqing 400044, China*

[3]*Theoretical Division and Center for Integrated Nanotechnologies (CINT-LANL), Los Alamos National Laboratory, Los Alamos, New Mexico 87545, USA*

[4]*Materials Department, University of California, Santa Barbara, California 93106, USA*

[5]*Stanford Synchrotron Radiation Lightsource, SLAC National Accelerator Laboratory, Menlo Park, California 94025, USA*

[6]*Department of Physics, Northeastern University, Boston, Massachusetts 02115, USA*

[7]*Quantum Materials and Sensing Institute, Northeastern University, Burlington, MA 01803, USA*

[#]These authors contributed equally to this work.

*Contact author. Email: jfhe@ustc.edu.cn

**Many emergent phenomena appear in doped Mott insulators near the insulator-to-metal transition. In high-temperature cuprate superconductors, superconductivity arises when antiferromagnetic (AFM) order is gradually suppressed by carrier doping, and a *d*-wave superconducting gap forms when an enigmatic nodal gap evolves into a point node. Here, we examine electron-doped $Sr_2IrO_4$, the 5*d*-electron counterpart of cuprates, using angle-resolved photoemission spectroscopy. At low doping levels, we observe the formation of electronic states near the Fermi level, accompanied by a gap at the AFM zone boundary, mimicking the AFM gap in electron-doped cuprates. With increasing doping, a distinct gap emerges along the (0,0)-($\pi$,$\pi$) nodal direction, paralleling that observed in hole-doped cuprates. This anomalous nodal gap persists after the collapse of the AFM gap and gradually decreases with further doping. It eventually vanishes into a point node of the reported**

**_d_-wave gap. These observations replicate the characteristic features in both electron- and hole-doped cuprates, indicating a unified route toward nodal metallicity in doped spin-1/2 AFM Mott insulators.**

A core challenge in doped Mott insulators is to understand the insulator-to-metal transition, where anomalous metallic states emerge from the insulating state. In high-temperature cuprate superconductors, continuous efforts have been dedicated to uncovering how nodal metallicity and *d*-wave superconductivity arise with doping [1–5]. On the electron-doped side, doping induced occupation of the conduction band enables a clear identification of the AFM gap in the electronic structure by photoemission measurements [6]. This gap opens at the antiferromagnetic zone boundary (AFMZB) in both nodal and anti-nodal regions. It decreases and vanishes with further electron doping [6–9]. On the hole-doped side, a pseudogap appears predominantly in the anti-nodal region and coexists with superconductivity over a wide doping range [2,5]. In the meantime, another enigmatic energy gap is observed along the nodal direction in the lightly doped regime [10–15], which collapses into a *d*-wave node with increased doping [2,5,13]. This nodal gap is not at the AFMZB, but whether it is related to the AFM order remains unclear. Despite the richness of these phenomena, the differences between electron- and hole-doped cuprates also make it complicated to differentiate the underlying fundamental physics from material-specific properties.

Electron-doped $Sr_2IrO_4$, a prototypical spin-orbit coupled Mott insulator [16], is theoretically described by the same Hubbard model as that of cuprates [17,18]. On one hand, it shares the same type of doped carriers with electron-doped cuprates. The on-site Coulomb repulsion $U$ is also moderate in both systems. On the other hand, there is a sign difference in the next-nearest hopping term (*t'*) of the one band Hubbard model between $Sr_2IrO_4$ and cuprates, which can be reversed by electron-hole conjugation [17]. Consequently, the low energy physics, if described by the Hubbard model, should be similar between electron-doped $Sr_2IrO_4$ and hole-doped cuprates [17,18]. As an electron-doped version to parallel hole-doped cuprates, electron-doped $Sr_2IrO_4$ not only provides a new window to examine the doping evolution of spin-1/2 AFM Mott insulators, but also offers a tantalizing perspective for understanding the phenomenology of a nodal metal [19-28]. This uniqueness makes it a valuable system for exploring the physics of the Hubbard model and gaining deeper insights into the mechanism that drives the exotic phenomena in cuprates.

In this paper, we report emergent phenomena associated with the insulator-to-metal transition in electron-doped $Sr_2IrO_4$. At low doping levels, angle-resolved photoemission spectroscopy (ARPES) measurements [see Supplementary Material (SM) Sec. S1 [29]] reveal the appearance of in-gap states, accompanied by an energy gap at the AFMZB, reminiscent of the AFM gap in electron-doped cuprates. With increasing doping, a distinct gap emerges at the Fermi level along the (0,0)-(π, π) nodal direction. This nodal gap is not located at the AFMZB in the momentum space, paralleling the mysterious nodal gap observed in lightly hole-doped cuprates [10-15]. With further doping, the AFM gap collapses, while the nodal gap gradually decreases and eventually vanishes into a point node, giving rise to nodal metallicity. These observations reproduce the characteristic features in both electron- and hole-doped cuprates and provide key insights into the formation of nodal metallicity in doped spin-1/2 AFM Mott insulators.

First, we have employed *in situ* potassium surface deposition on pristine $Sr_2IrO_4$ to gradually increase the electron concentration and continuously track the doping evolution of the electronic structure. Figure 1 shows the detailed doping dependence of the ARPES spectra measured across the nodal (Figs. 1a-c) and anti-nodal (Figs. 1d-f) regions. In the lightly doped regime, an overall downward shift of the lower Hubbard band (LHB) is observed with doping (Figs. 1b,e, see SM Secs. S2 and S3 for details [29]). Upon further electron doping, in-gap states begin to appear near the Fermi level ($E_F$), and the electron spectral weight is transferred from the LHB to these in-gap states (Figs. 1b,c,e,f). Meanwhile, the remnant LHB moves upward in energy, distinct from the initial downward shift in the lightly doped regime. These observations reveal a continuous evolution in electron-doped $Sr_2IrO_4$: initial electron filling induces an upward shift of the chemical potential, giving rise to a relative downward shift of the LHB; further doping creates in-gap states and decreases the Mott gap via an enhanced screening effect, leading to an upward shift of the LHB (Figs. 1b,e).

After understanding the overall doping evolution on a large energy scale, we focus on the in-gap states near $E_F$ (Fig. 2). Measurements on a bulk electron-doped sample [$(Sr_{1-x}La_x)_2IrO_4$, x∼0.029] clearly reveal an electron-like band and a hole-like band formed by the in-gap states (Fig. 2b, marked as "IGSs"). These two bands are separated by an energy gap at (π/2, π/2) where they intersect the AFMZB [Figs. 2b (ii and iii)]. To better understand this gap at the AFMZB, continuous surface doping on the parent compound (Fig. 2a) has been carried out. As shown in Figs. 2c (ii-iv), the in-gap states are initially gapped from $E_F$, forming a hole-like band centered at (π/2, π/2). The magnitude of the gap decreases with doping, as the top of the hole-like band moves towards $E_F$. With further doping, an electron-like band starts to appear below $E_F$ [Figs. 2c(v-viii)], and the energy gap at the AFMZB vanishes when the bottom of the electron-like band touches the top of the hole-like band. This doping evolution (Figs. 2b,c) is distinct from electrons filling a rigid band. It is consistent with an energy gap at the AFMZB, whose magnitude decreases with electron doping (Fig. 2e). We note that the hole-like band of the in-gap states would also intersect the AFMZB at $(0, \pi)$ in the anti-nodal region (Fig. 1f). However, the crossing point and possible gap opening are above $E_F$, which cannot be probed by photoemission measurements (see SM Sec. S4 [29]).

In the doping regime, where the electron-like band bottom is below $E_F$, the energy gap at AFMZB, characterized by the energy difference between the electron-like band bottom and hole-like band top, would not appear at $E_F$. However, a careful examination suggests a striking spectral weight suppression at the underlying Fermi momentum $k_F$, where the electron-like band should intersect $E_F$ [Fig. 2c (vi)]. This anomalous gap becomes more evident on a bulk electron-doped sample [$(Sr_{1-x}La_x)_2IrO_4$, x~0.041] with a freshly cleaved surface (Fig.3). The gap at the AFMZB is already closed in this sample, but a spectral weight suppression is observed at $E_F$, indicating the existence of another energy gap [Fig. 3a (i)]. This gap is not at the AFMZB but on the underlying Fermi surface along the zone diagonal direction (marked as underlying Fermi momentum $k_F$), paralleling the nodal gap reported in hole-doped cuprates [10–15]. Therefore, we label this new gap as nodal gap, hereafter. The detailed doping dependence of the nodal gap is obtained by continuous potassium surface deposition on the same sample (Fig. 3a). The increased electron doping can be characterized by the increased distance between the Fermi momenta on the main band and folded band (Fig. 3a). The magnitude of the gap is quantified by the energy distribution curve (EDC) at $k_F$ of the main band (Figs. 3b,c), and its doping evolution is summarized in Fig. 3d (see SM Secs. S5-S7 for details [29]). It is clear that the nodal gap decreases as a function of electron doping and vanishes at a critical doping level (Figs. 3b-d). In the meantime, a quasiparticle peak starts to emerge and sharpens with electron doping (Figs. 3b,c,e). This trend continues even after the collapse of the nodal gap.

The above observations collectively depict an integrated picture of the doping evolution in electron-doped $Sr_2IrO_4$ (Fig. 4). Initial doping induces in-gap states and the electron density of states is transferred from the Hubbard band to the in-gap states. At low doping levels, these in-gap states are split by an energy gap at the AFMZB (Fig. 4a). This gap vanishes at a doping level where the AFM order disappears, as observed in magnetization, neutron and x-ray scattering measurements [24–26] (Fig. 4b). Strikingly, another nodal gap also develops and persists even after the collapse of the gap at the AFMZB (Figs. 4b,c). It gradually decreases with further electron doping and disappears at a doping level close to where the *d*- wave gap and anti-nodal pseudogap are reported in this system [19,20,27]. The resulting electronic phase diagram, as summarized in Figs. 4b and 4c, illustrates how the nodal metallicity emerges from the Mott insulating state of $Sr_2IrO_4$.

We now discuss the implications of these observations. First, the doping-induced in-gap states and the existence of remnant LHB over a wide doping range (Figs. 1b,e) are well aligned with previous reports in both electron- and hole-doped cuprates [2,5,6]. Second, the observed energy gap at the AFMZB phenomenologically parallels the AFM gap observed in electron-doped cuprates [6–9]. In electron-doped $Sr_2IrO_4$, both structural octahedral rotation and AFM order give rise to the same scattering wave vector (π, π) [28], which could reconstruct the original band into electron-like and hole-like bands with a gap opening at the AFMZB. However, the gap is absent in the bulk electron-doped sample $(Sr_{1-x}La_x)_2IrO_4$, x~0.041, in which the structural octahedral rotation still persists [28]. This observation makes the structural origin unlikely. Considering the observation that the gap at the AFMZB vanishes at a doping level where the AFM order also disappears (Fig. 4b), it is more compelling to attribute the gap to a magnetic origin. This is also supported by our doping-dependent model calculations of an AFM order in $Sr_2IrO_4$. As shown in Fig. 2d, an energy gap appears at the AFMZB and vanishes with doping, mimicking the experimental observations (Fig. 2c).

The mysterious nodal gap in electron-doped $Sr_2IrO_4$ shares striking similarities with that reported in hole-doped cuprates [10–15], including its location in the Brillouin zone, the doping dependence of the gap magnitude, and the evolution of the associated quasiparticle spectral weight (Fig. 3 and Fig. 4c). This anomalous nodal gap in electron-doped $Sr_2IrO_4$ is not directly driven by the AFM order, as it exists when the AFM long range order is suppressed. A Coulomb gap has been suggested in lightly doped cuprates, where Coulomb disorder effects can localize the electronic states near $E_F$ and create an energy gap [36]. In the current case, the Coulomb gap scenario may explain the collapse of nodal gap with doping due to an enhanced screening of the Coulomb interaction (see SM Secs. S1 and S8 for simulations [29]). However, this scenario cannot explain the observed quasiparticle peaks and fails to describe the evolution of the spectral line-shape with doping (Figs. 3b,c). Polaron effects have been discussed in hole-doped cuprates to describe the emergence of quasiparticle peak with doping, but they cannot explain the energy gap [13,37]. We note that an incommensurate spin density wave state has been reported [26] in electron-doped $Sr_2IrO_4$ at a doping level similar to where the nodal gap is identified. In cuprates, it has been proposed that the spin density wave order may give rise to an energy gap [15]. But whether it can account for the nodal gap remains to be explored. A scenario of topological superconductor has also been proposed theoretically to understand the nodal gap in cuprates [38]. It could be relevant to the nodal gap in $Sr_2IrO_4$, if the reported low-temperature *d*-wave gap represents superconductivity [19,39].

While the origin of the nodal gap remains to be fully understood, the striking parallel of the overall doping evolution between $Sr_2IrO_4$ and cuprates indicates that the emergence of in-gap states, the collapse of the AFM gap, and the development of a nodal gap are not exclusive to either electron- or hole-doped cuprates. Instead, they may represent generic phenomena in doped spin-1/2 AFM Mott insulators. It is worth noting, however, that an earlier study on hole-doped $Sr_2Ir_{1-x}T_xO_4$ (T = Rh, Ru) has shown that substituting Ir by 4*d* elements can modify the spin-orbit coupling (SOC) strength, thereby influencing the insulator-to-metal transition [40]. In that case, the doping effect is intrinsically linked to changes in SOC. In our study, continuous electron doping is achieved via *in situ* deposition of potassium and La substitution of Sr. The Ir sublattice remains intact and no substantial change of SOC is observed.

Our results, when combined with previous reports of pseudogap and *d*-wave gap in $Sr_2IrO_4$ [19,20,27], point to a potentially unified doping pathway from the Mott insulating state to a nodal metallic state (Fig. 4c). Although superconductivity has yet to be experimentally observed in doped $Sr_2IrO_4$, the universality we highlight primarily pertains to the normal state properties leading toward the nodal metal, independent of whether superconductivity eventually emerges. These findings motivate future studies to explore the interplay between universal doping evolution, nodal metallicity, and the potential realization of high-$T_c$ superconductivity in doped spin-1/2 Mott systems.

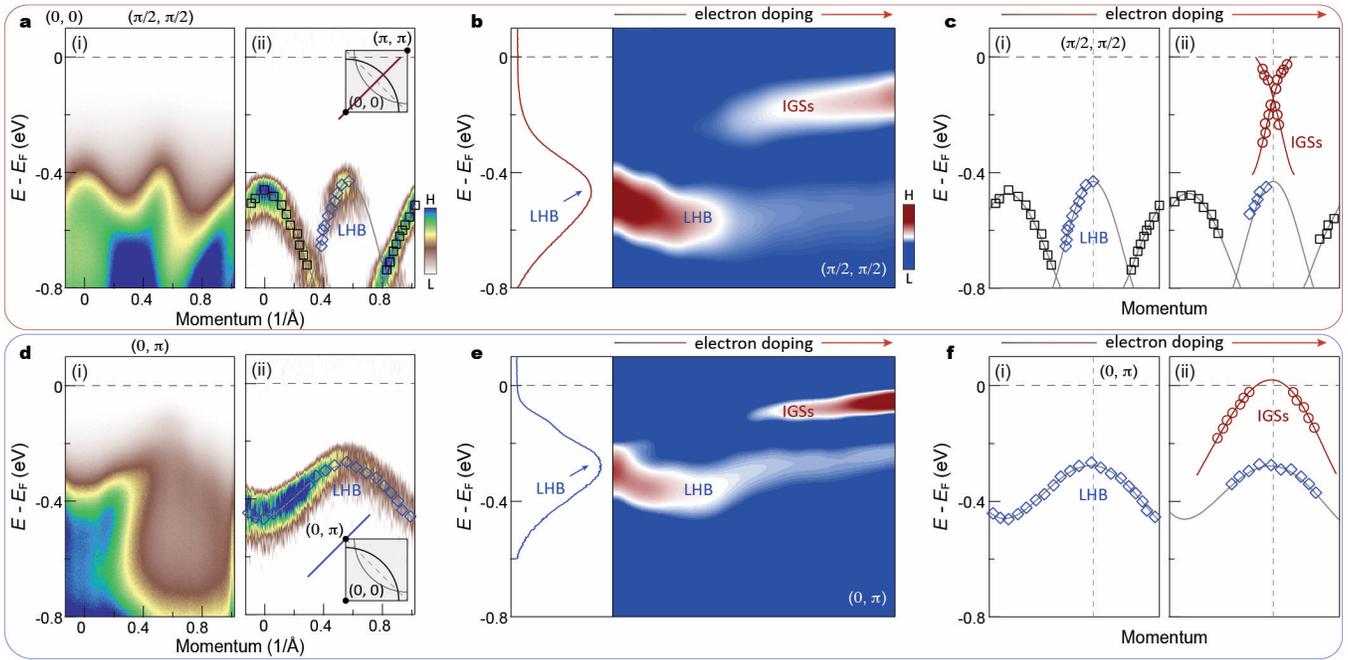

**Fig. 1. Overall doping dependence of the electronic structure in electron doped $Sr_2IrO_4$. a,** ARPES spectrum (i) and its second derivative as a function of energy (ii), taken along the (0, 0) - (π, π) nodal direction as indicated by the red line in the inset of (ii). **b,** Energy distribution curve (EDC) at the (π/2, π/2) nodal region of the parent compound (red curve, left panel) and its continuous evolution as a function of electron doping (image, right panel) (see Fig. S2 for details [29]). **c,** Extracted band dispersion along the (0, 0) - (π, π) nodal direction for the parent (i) and the heavily electron-doped (ii) compound. **d-f,** Same as (**a-c**), but for the (0, π) antinodal region.

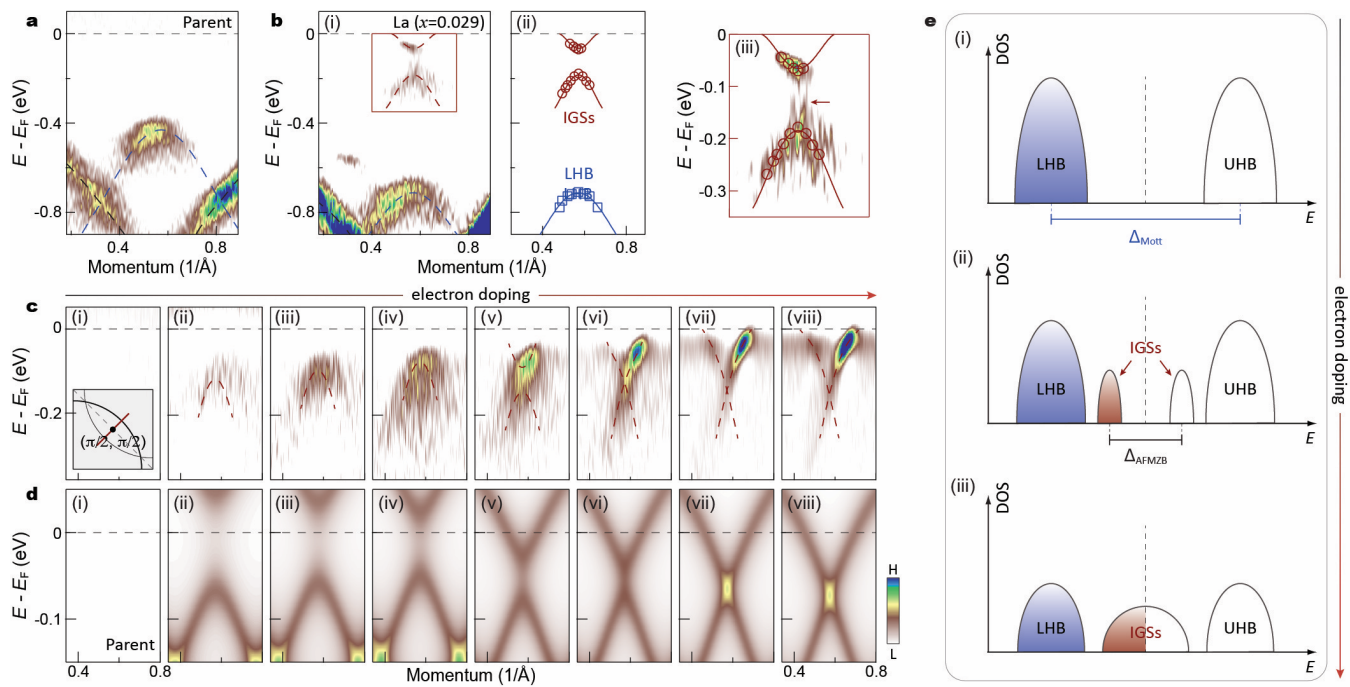

**Fig. 2. Evolution of the in-gap states as a function of electron doping. a,** Absence of IGSs in the parent compound. **b,** IGSs in the La-doped (x=0.029) sample (i) and the extracted band dispersion (ii). The electron-like band and hole-like band formed by the IGSs are shown in an expanded scale in (iii). **c,** Doping evolution of the IGSs measured along the nodal direction (inset of i) at 90 *K*. Second derivative spectra as a function of energy are shown in (**a**-**c**) to enhance the fine features. **d,** Simulated spectra along the nodal direction. **e,** Schematic diagram of the electronic states as a function of electron doping.

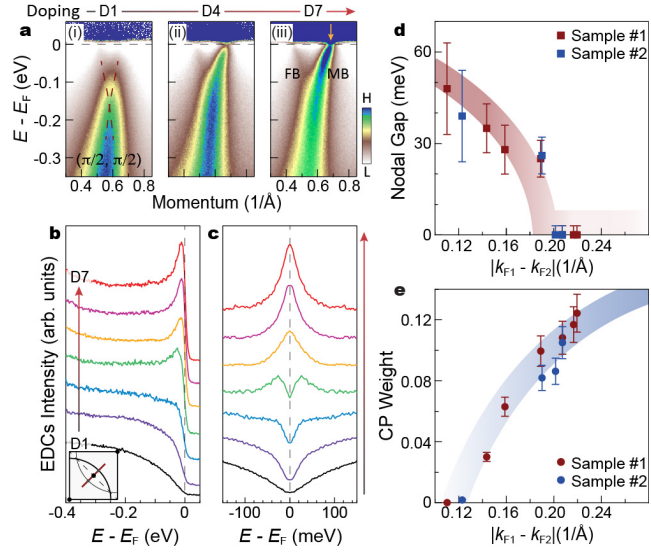

**Fig. 3. Observation of an anomalous nodal gap. a,** Fermi-Dirac-divided ARPES spectra along the nodal direction as a function of doping, measured at 30 K. The main band (MB) and folded band (FB) are marked [21,27]. D1-D7 indicate the doping sequences. D1 represents the measurement on the pristine $(Sr_{1-x}La_x)_2IrO_4$, x~0.041 sample without potassium deposition. **b-c,** Doping dependence of the raw EDC (**b**) and symmetrized EDC (**c**) at the $k_F$ of the MB [indicated by the orange arrow in [a(iii)]. **d-e,** Doping evolution of the nodal gap (**d**) and the spectral weight of the quasiparticle coherence peak (CP) (**e**). See SM Sec. S5 [29] for details.

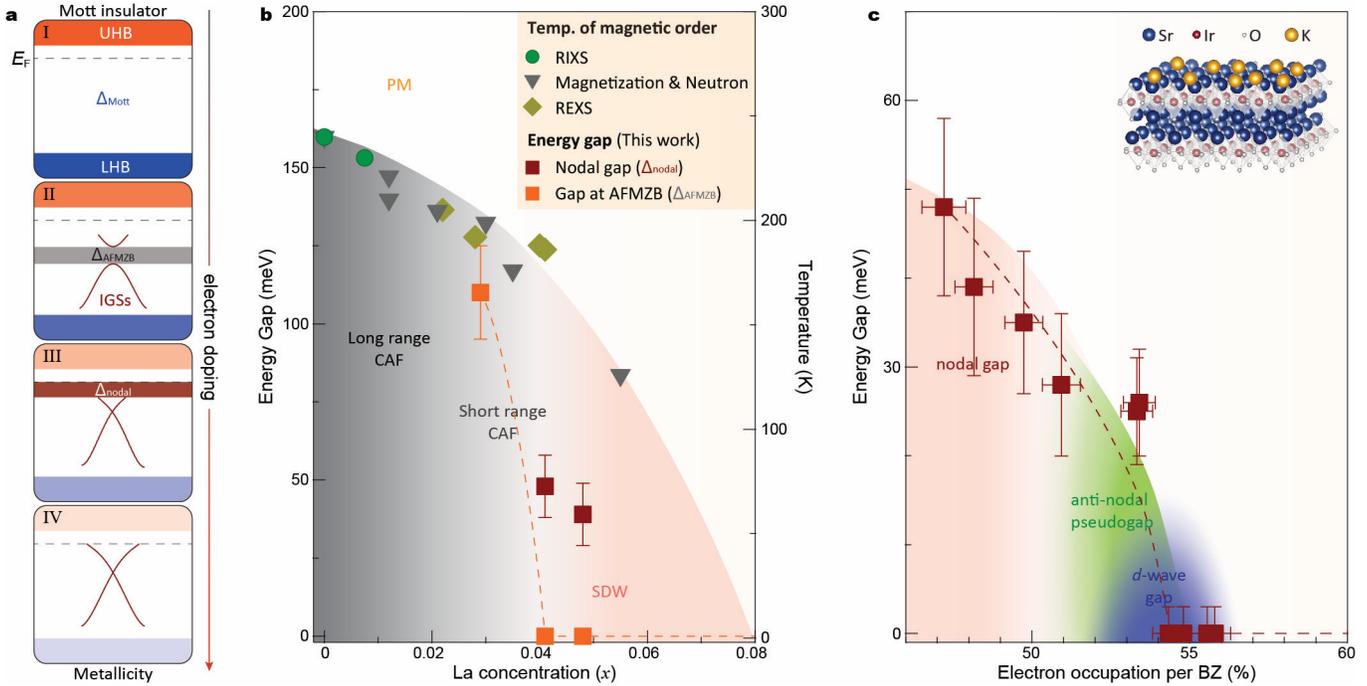

**Fig. 4. Electronic phase diagram of electron doped $Sr_2IrO_4$. a,** Schematic of the electronic structure as a function of electron doping. **b,** Phase diagram of bulk electron doped $(Sr_{1-x}La_x)_2IrO_4$. The evolution of the magnetic orders and the spin-density wave order are from reported results [24-26]. **c,** Extended phase diagram beyond the bulk electron doping limit. Further electron doping is realized by *in situ* potassium deposition on $(Sr_{1-x}La_x)_2IrO_4$ samples. Green and blue shaded regions in (**c**) are schematic of the anti-nodal pseudogap [20,27] and *d*-wave gap [19], respectively. The carrier concentrations of the surface doped samples are shown by the ratio between the area of the occupied underlying Fermi surface and that of the Brillouin zone [20].

13